# Data-mining the Foundational Patents of Photovoltaic Materials:

## An application of Patent Citation Spectroscopy

Jordan A. Comins [a] and Loet Leydesdorff [b]*

**Key messages:**

Keeping abreast of most technologically important patents is also a challenging task for patent granting organizations and patenting inventors and firms. As a consequence, patent offices have invested in developing novel automated approaches for identifying landmark patents in technology areas. Patent Citation Spectroscopy (PCS) provides a web-application to answer this question within minutes.

Patents branch out in tree-like structures along trajectories. The historical root, or seminal, patent can be followed using sequences of patent citations. The algorithmic method of PCS presented in this study provides a solution to the problem where to begin the analysis of a technological development. PCS enables the user to retrieve the fundamental patent in any technological domain using a topical search. This application thus orients the user strategically.

The online data-mining method of PCS is based on Reference Publication Year Spectroscopy (RPYS) technique, a methodology developed for use on academic literature. However, PCS includes additional normalization calculations to disentangle citation outliers based on the outstanding performance of a single document as compared to a group of documents.

To illustrate the value of PCS, we provide the results of a search for the seminal patents of the nine CPC subclasses pertaining to photovoltaic solar cells, a key area of technological innovation. Research and development (R&D) in photovoltaic devices continues to yield greater efficiencies, offering the potential to lower the cost of solar energy (Chu et al, 2016). As these advances in solar technology become primed for penetrating the global energy system, an understanding of the key patents and inventors in photovoltaic materials will assist decision-makers in understanding the R&D landscape (Polman et al, 2016). We demonstrate that such searches are easily completed via PCS in each of the nine CPC subclasses. Searches of scholarly article databases validated the results obtained through PCS in five of the nine classes.

[a] Social and Behavioral Sciences Department, The MITRE Corporation, McLean, VA, United States; jcomins@gmail.com ; * corresponding author
[b] Amsterdam School of Communication Research (ASCoR), University of Amsterdam, PO Box 15793, 1001 NG Amsterdam, The Netherlands; loet@leydesdorff.net.

# 1. Introduction

In his presidential address to the American Economic Association entitled "Productivity, R&D, and the Data Constraint", Grilliches (1994, p. 14) formulated as follows:

> Our measurement frameworks are not set up to record detailed origin and destination data for commodity flows, much less so for information flows. We do have now a new tool for studying some of this: citations to patents and the scientific literature (see e.g., Jaffe et al., 1993), but anyone currently active in the e-mail revolution and participating in the conferences and workshops circuit knows how small this tip is relative to the informal-communications iceberg itself.

In this study, we report on a routine (Patent Citation *Specroscopy* or *PCS*) for using citations among patents to trace the foundational patent so that one can more easily reconstruct the trajectory of a technology from its origin to the present time. Hitherto, subject matter experts have to review patents and patent applications and maintain an awareness of the most technologically important patents. This has remained a time-consuming practice which presents several obstacles including difficultly with reliability and replication, and dependence on the availability of experts (Cockburn et al., 2002).

The problem is not only one of sufficient (wo)man power and skills. The huge database is not easily accessible for retrieval and reconstruction. Jensen & Murray (2005), for example, argue that the impact of gene patents on downstream research and innovation are unknown, in part because of a lack of empirical data on the extent and nature of gene patenting. The intellectual property rights for some genes can become highly fragmented between many owners, which suggests that downstream innovators may face considerable costs to gain access to gene-oriented technologies. Konski and Spielthenner (2009) developed a landscape analysis of stem-cell patents using a clustering algorithm based on network analysis enabling the user to find "bridging" patents between technological developments.

With the support of the Office of the Chief Economist in the US Patent & Trademark Office (USPTO), PatentsView was launched in 2015 as a new patent data visualization and analysis platform intended to increase the value, utility, and transparency of US patent data. The PatentsView platform is built on USPTO's regularly updated database that longitudinally links inventors, their organizations, locations, and overall patenting activity. PatentsView delivers US patent data in ways that enable this data to be fully discoverable and exploitable by various end users. Our algorithmic method for Patent Citation Spectroscopy (PCS) exploits PatentsView data and enables the user to identify landmark patents interactively via a web-application (url: http://www.leydesdorff.net/comins/pcs/index.html

## 2. Retrieval and Disclosure: Patents as Indicators

Beyond their critical role in industry, patents are indicators of inventions and thus can be expected to carry information about technological progress (O'Donoghue et al., 1998; Jaffe, Trajtenberg, & Henderson, 1993; Harhoff et al., 1999; Artz et al., 2010; Graevenitz et al., 2013; Comins, 2015). Patents provide a unique window on knowledge-based economies (Jaffe & Trajtenberg, 2002) and can serve as both an indicator of industrial activity and output of academia (Shelton & Leydesdorff, 2012). The United States Patent and Trademark Office (USPTO) under the Department of Commerce plays a vital role in relating university and industry in the American innovation system by registering and extending legal protection over inventions. In exchange for detailed public disclosure of a technical invention, the patent assignee, the legal entity to which intellectual property rights are assigned, is entitled to a monopoly over the patent's claims.

Let us as an example demonstrate the effectiveness of PCS for the retrieval by conducting an analysis of the seminal patents for the material technologies underlying photovoltaic cells. As R&D in photovoltaic materials matures, increases in energy efficiency and decreases in production costs could enable a significant impact on the global energy sector (Polman et al, 2016; Chu et al., 2016). Intellectually, this study follows up on Leydesdorff, Alkemade, Heimeriks, & Hoekstra's (2015) study of the innovation dynamics of photovoltaic cells providing an animation of geographical diffusion at http://leydesdorff.net/photovoltaic/cuinse2/index.html. (See also at http://www.leydesdorff.net/software/patentmaps/dynamic/ for instruction.) In that study, however, we focused only on "dye sensititzed solar cells" ($CuInSe_2$-based cells), its geographical diffusion, and technological branching from the perspective of technology studies and regional economics. In this study, we do not follow the time axis, but look back in order to a starting point for the evolving technology. Furthermore, we extend the analysis to the nine classifications recently added to the patent classifications for photovoltaic cells (Table 1). To do so, we leverage the taxonomy of the recently renewed patent classification system, known as the Cooperative Patent Classification (CPC).

While there are numerous patent classification systems, among the most widely-used in patent studies are hitherto the United States Patent Classification (USPC) system, which comprises more than 160,000 classes and subclasses of patent functions (USPTO, 2008), its European counterpart (ECLA of the European Patent Office EPO) and the International Patent Classification (IPC) system, a hierarchical system managed by the World Intellectual Property Organization (WIPO) consisting of more than 70,000 classifications of technical fields (WIPO, 2014). In 2013, the USPTO and the EPO adopted a new classification system for patents that will ultimately replace both the USPC and IPC. The CPC system of these two large agencies provides a tree-like hierarchy consisting of 5-levels of depth and more than 250,000 classifications at the level of the leaf node and is currently in use for patents filed through the USPTO as well as EPO.

Furthermore, CPC adds to the previous systems by the introduction of the Y-class of patents representing newly emerging technologies across sectors. The new classes are backtracked into the previous system. Currently, there are nine CPC classifications that describe material photovoltaic technologies:

Table 1: Nine classes of photovoltaic cells in CPC

| | |
|---|---|
| Y02E 10/541 . . . | CuInSe2 material PV cells |
| Y02E 10/542 . . . | Dye sensitized solar cells |
| Y02E 10/543 . . . | Solar cells from Group II-VI materials |
| Y02E 10/544 . . . | Solar cells from Group III-V materials |
| Y02E 10/545 . . . | Microcrystalline silicon PV cells |
| Y02E 10/546 . . . | Polycrystalline silicon PV cells |
| Y02E 10/547 . . . | Monocrystalline silicon PV cells |
| Y02E 10/548 . . . | Amorphous silicon PV cells |
| Y02E 10/549 . . . | organic PV cells |

We extend our understanding of the performance of PCS by applying the methodology for each of these classifications using the advanced search capability of the online tools of PatentsView and PCS. Below we first briefly review the PCS methodology and tool, and then describe our findings pertaining to the landmark patents underlying photovoltaic material technologies.

**2. Patent Citation Spectroscopy**

PCS is a data mining method that operates over the cited references within sets of patents. The goal is to generate a historical assessment of the most impactful patents within technological areas. The underlying PCS computation is based on a similar data mining methodology developed for use on academic literature, known as Reference Publication Year

Spectroscopy (RPYS) technique (Marx, Bornmann, Barth, & Leydesdorff, 2014). This method involves aggregating the cited references across a set of retrieved documents and organizing these cited references by their publication year. For each cited reference year, the total number of references is calculated. Next, data is de-trended by taking the absolute deviation of the number of cited references for a given year from the 5-year median. As specifically applied to patents, this is represented by the equation:

$$f(t) = C_t - med(C_{t-2}, C_{t-1}, C_t, C_{t+1}, C_{t+2}), \qquad (1)$$

where *C* represents the total sum of citations to patents granted in year *t* and *med* represents the median. These steps do not deviate from RPYS in calculation (though RPYS was never applied to patents). However, this de-trending function only considers the aggregated cited reference activity over time. This creates a challenge in identifying seminal works because interesting outliers resulting from the de-trending equation could result from either a large surge in the influence of a single document (i.e., what we might consider a seminal work) or based on several slightly influential documents occurring in the same year. As such, PCS includes an additional normalization calculation to disentangle outliers based on the outstanding performance of a single document as compared to a group of documents:

$$PCS(t) = f(t) \cdot \frac{\text{Count of References to Most Referenced Patent in Year } t}{C_t} \qquad (2)$$

This step multiples the results from equation (1) based on the percentage of all references from that year attributable to the most referenced patent.

**3. Applying Patent Citation Spectroscopy to Material Photovoltaic Technologies**

At present, PCS can be applied to granted US patents using a web-application produced by Comins et al. (*under* review; http://www.leydesdorff.net/comins/pcs/index.html). The web-application leverages the application programming interface (API) to the public data platform PatentsView, which is a supported by the USPTO Chief Economist. Users can search for patents using either keyword phrases (e.g., "photovoltaic cells") or more advanced searches. These advanced searches follow the conventions described by the data-provider (PatentsView) documentation. Among other things, advanced search queries enable users to apply PCS to patents based on their Cooperative Patent Classification.

Using the PCS web-application, we conducted a search for the seminal patents of the nine CPC subclasses pertaining to photovoltaic solar cells. Here, we walk through the analytic routine for a single case (CPC subclass Y02E 10/541: $CuInSe_2$ material PV cells). In this case, an advanced search was conducted in the PCS-application using the following query: ADVANCED={"cpc_subgroup_id":"Y02E10\/541"}. This search retrieved metadata on 962 granted US patents and analyzed a total of 3,502 unique patent references. The application yields a visualization of the PCS algorithm output as well as the method's most likely seminal patent (see Figure 1). In the case of CPC subclass Y02E 10/541, the resulting seminal patent is US4335266: "Methods for forming thin-film heterojunction solar cells from I-III-IV$_2$" by Reid Mickelsen and Wen Chen.

To validate the results of the algorithm, we conduct a search for scholarly articles citing patent US4335266 as the underlying invention of $CuInSe_2$ material PV cells. In this instance, an article appearing in *Materials Science Forum* states "…in 1980, Boeing Aerospace demonstrated, for the first time, the milestone of 10 % small-area cell efficiency in the form of thin-film solar cells with a CuInSe2 alloy system, in which they successfully invented how to prepare the p-type absorbers known as so-called 'bilayer' process [Mickelsen and Chen, US4335266]." Such articles provide corroborating evidence as the performance of PCS (cf. Leydesdorff, Alkemade, Heimeriks, & Hoekstra, 2015).

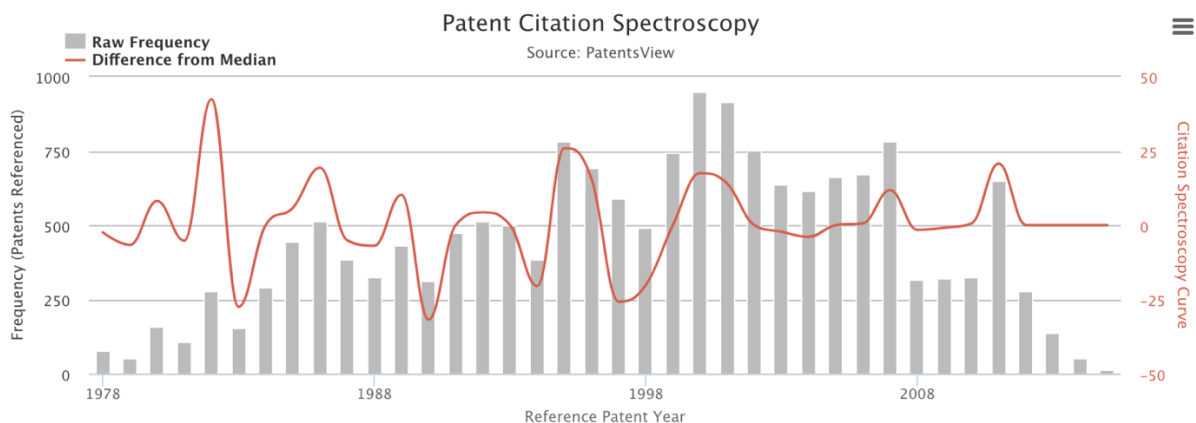

**Figure 1**. The PCS-derived foundational patent for CPC Subclass Y02E 10/541: $CuInSe_2$ material PV cells is US4335266: "Methods for forming thin-film heterojunction solar cells from I-III-IV$_2$" by Reid Mickelsen and Wen Chen.

This foundational patent was granted in 1982 and is cited 151 times since then in other USPTO patents. Figure 2 shows the time series of the patents building on US4335266 and broken down for country names. The number of co-inventors is 351, of which 56 from Japan, 10 from Taiwan, and 273 from the USA. However, 82% of the applicants are American. The Japanese and Taiwanese efforts during the period 1995-2005 were perhaps too early. The main applications followed after three decades in the US during the years 2010-2015.

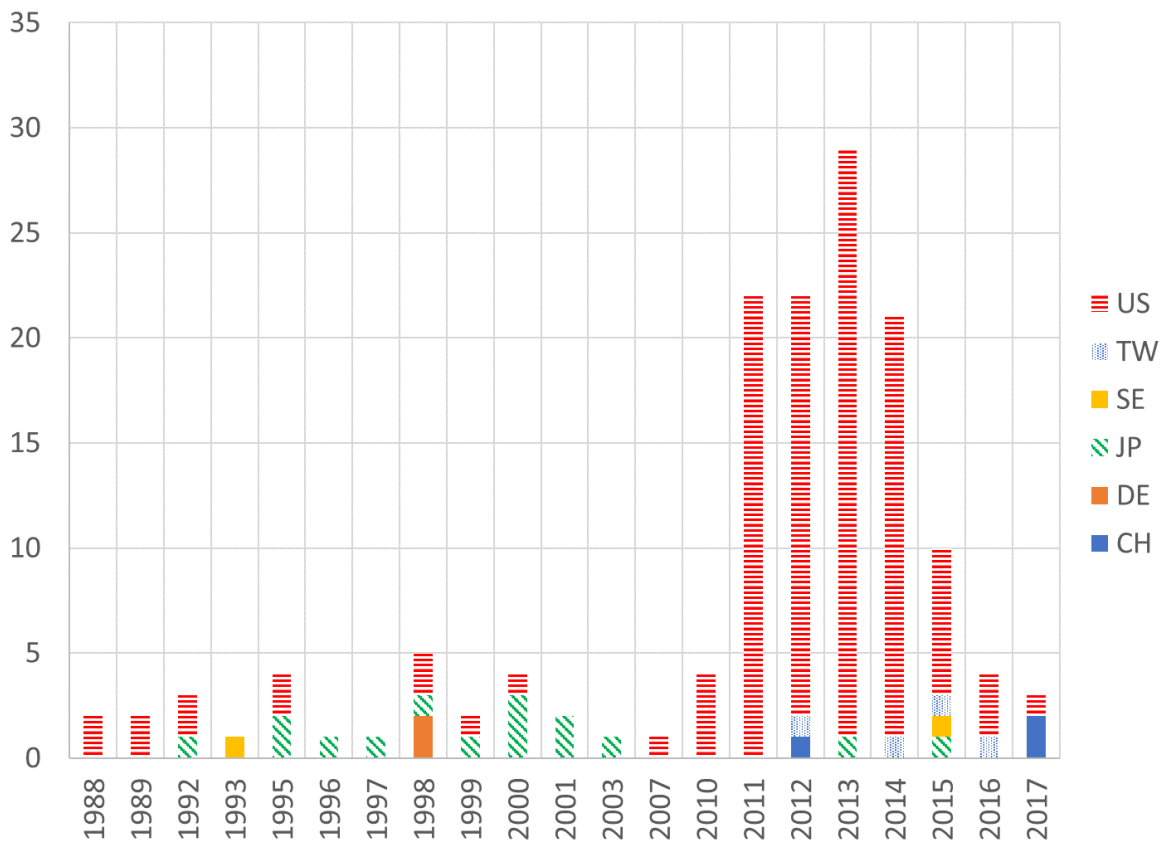

**Figure 2**: Geographical spread over time of the 151 US patents citing the foundational patent of CuInSe2 material PV cells.

In sum, the technique enables us to find the foundational patent and to pursue the analysis from there. A summary of results for all nine photovoltaic subclasses examined in our study are detailed in Table 1. Once a seminal patent for a given topic was identified, a traditional search of scholarly literature was performed to identify corroborating evidence of the patent's influence. In the case of 5 of the 9 subgroups analyzed, we found corroborating scholarly evidence for the seminal patents identified by the PCS algorithm.

| Patent Topic | CPC Subgroup | PCS Identified Seminal Patent | Corroborating Evidence |
| --- | --- | --- | --- |
| CuInSe2 material PV cells | Y02E10\V541 | US4335266 | Kushiya et al., 2012 |
| Dye sensitized solar cells | Y02E10\V542 | US4927721 | Longo & Paoli, 2003 |
| Solar cells from Group II-VI materials | Y02E10\V543 | US5536333 | Cheese et al., 2016 |
| Solar cells from Group III-V materials | Y02E10\V544 | US6252287 | Takamoto et al., 2005 |
| Microcrystalline silicon PV cells | Y02E10\V545 | US5677236 | *none* |
| Polycrystalline silicon PV cells | Y02E10\V546 | US5227329 | *none* |
| Monocrystalline silicon PV cells | Y02E10\V547 | US5053083 | *none* |
| Amorphous silicon PV cells | Y02E10\V548 | US4109271 | *none* |
| Organic PV cells | Y02E10\V549 | US4539507 | Kiy, 2002 |

**Table 2**. Summary results from our application of the PCS algorithm to 9 CPC subclasses related to photovoltaic technology. Table provide corroborating evidence, where available, for the PCS tool.

## 4. Summary and Conclusions

We used Patent Citation Spectroscopy—originally developed as Reference Publication Year Spectroscopy for studying landmarks and milestones in scientific literature (Comins & Leydesdorff, 2017; Thor, Marx, Leydesdorff, & Bornmann, 2016)—to patent literature classified into the nine Y-subclasses of CPC that describe material photovoltaic technologies. In five of the nine cases, we found corroborating evidence for the foundational character of the patent indicated by the routine.

The possible applications of PCS are numerous. In a scholarly context, one can be interested in the reconstruction of the main path of patent citations (Liu & Lu, 2012). Patents branch out in tree-like structures along trajectories. The root patent can be followed historically using sequences of patent citations. The algorithmic method for Patent Citation Spectroscopy (PCS) presented in this study provides solution to the problem where to begin the analysis of a technogical development. PCS enables the user to retrieve the fundamental patent in any technological domain using a topical search. This application thus orients the user strategically (Rotolo et al., 2017).

For this study we extended the routine with the option to use the advanced search queries at PatentsView. On the basis of two normalizations of the longitudinal distribution of the publication years of the patents cited by the retrieved patents, the routine (at http://www.leydesdorff.net/comins/pcs/index.html) provides a best guess of the foundational patent for the subject specified in the string. It seems to us that the successful application in five of the nine cases and the previous results in the case of biomedical patents reported by Comins et al. (2017) provide some confidence that this indicator of fundamental patents has potential. However, the normalizations may have to be refined based on further analysis of successful and unsuccessful applications.

Follow-up studies could combine the results of PCS with the longitudinal animations demonstrated at http://leydesdorff.net/photovoltaic/cuinse2/index.htm , but since further developed into a stand-alone tool *PatViz*. (The latest release of PatViz can be downloaded from https://github.com/Data2Semantics/PatViz/releases or http://www.leydesdorff.net/photovoltaic/patviz/ ) for installation on one's own machine. One can upload one's own data to this routine. For example, one can first retrieve the source of a technology using PCS and then follow the citations.

**References**

Cheese, E., Mapes, M. K., Turo, K. M., & Jones-Albertus, R. (2016, June). US department of energy photovoltaics research evaluation and assessment. In *Photovoltaic Specialists Conference (PVSC), 2016 IEEE 43rd* (pp. 3475-3480). IEEE.

Chu, S., Cui, Y., & Liu, N. (2017). The path towards sustainable energy. *Nature materials*, *16*(1), 16.

Comins, J. A., & Leydesdorff, L. (2017). Citation algorithms for identifying research milestones driving biomedical innovation. *Scientometrics, 110*(3), 1495-1504.

Comins, J. A., Carmack, S. A., & Leydesdorff, L. (under review). Patent Citation Spectroscopy (PCS): Algorithmic retrieval of landmark patents. *arXiv preprint arXiv:1710.03349*.

Fire, A., Xu, S. Q., Montgomery, M. K., Kostas, S. A., Driver, S. E., & Mello, C. C. (1998). Potent and specific genetic interference by double-stranded RNA in *Caenorhabditis elegans*. *Nature, 391*(6669), 806-811.

Griliches, Z. (1994). Productivity, R&D and the Data constraint. *American Economic Review, 84*(1), 1-23.

Jaffe, A. B., & Trajtenberg, M. (2002). *Patents, Citations, and Innovations: A Window on the Knowledge Economy*. Cambridge, MA/London: MIT Press.

Jaffe, A. B., Trajtenberg, M., & Henderson, R. (1993). Geographic localization of knowledge spillovers as evidenced by patent citations. *the Quarterly journal of Economics, 108*(3), 577-598.

Jensen, K., & Murray, F. (2005). Intellectual property landscape of the human genome. *Science, 310*(5746), 239-240.

Kiy, M. (2002). *Charge injection and transport in organic semiconductors* (Doctoral dissertation).

Konski, A. F., & Spielthenner, D. J. (2009). Stem cell patents: a landscape analysis. *Nature biotechnology, 27*(8), 722.

Kushiya, K., Sugimoto, H., Chiba, Y., Tanaka, Y., & Hakuma, H. (2012). Change of the Characterization Techniques as Progress of CuInSe2-based Thin-Film PV Technology. In *Materials Science Forum* (Vol. 725, pp. 165-170). Trans Tech Publications.

Leydesdorff, L., Alkemade, F., Heimeriks, G., & Hoekstra, R. (2015). Patents as instruments for exploring innovation dynamics: geographic and technological perspectives on "photovoltaic cells". *Scientometrics, 102*(1), 629-651. doi: 10.1007/s11192-014-1447-8

Liu, J. S., & Lu, L. Y. (2012). An integrated approach for main path analysis: Development of the Hirsch index as an example. *Journal of the American Society for Information Science and Technology, 63*(3), 528-542.

Longo, C., & De Paoli, M. A. (2003). Dye-sensitized solar cells: a successful combination of materials. *Journal of the Brazilian Chemical Society*, *14*(6), 898-901.

Marx, W., Bornmann, L., Barth, A., & Leydesdorff, L. (2014). Detecting the historical roots of research fields by reference publication year spectroscopy (RPYS). *Journal of the Association for Information Science and Technology, 65*(4), 751-764.

Polman, A., Knight, M., Garnett, E. C., Ehrler, B., & Sinke, W. C. (2016). Photovoltaic materials: Present efficiencies and future challenges. *Science*, *352*(6283).

Rotolo, D., Rafols, I., Hopkins, M. M., & Leydesdorff, L. (2017). Strategic intelligence on emerging technologies: Scientometric overlay mapping. *Journal of the Association for Information Science and Technology, 68*(1), 214-233. doi: 10.1002/asi.23631

Shelton, R. D., & Leydesdorff, L. (2012). Publish or Patent: Bibliometric evidence for empirical trade-offs in national funding strategies. *Journal of the American Society for Information Science and Technology, 63*(3), 498-511.

Takamoto, T., Agui, T., Washio, H., Takahashi, N., Nakamura, K., Anzawa, O., ... & Yamaguchi, M. (2005, January). Future development of InGaP/(In) GaAs based multijunction solar cells. In *Photovoltaic Specialists Conference, 2005. Conference Record of the Thirty-first IEEE* (pp. 519-524). IEEE.

Thor, A., Marx, W., Leydesdorff, L., & Bornmann, L. (2016). Introducing CitedReferencesExplorer : A program for Reference Publication Year Spectroscopy with Cited References Disambiguation. *Journal of Informetrics, 10*(2), 503-515. doi: 10.1016/j.joi.2016.02.005